\documentclass[11pt,a4wide]{article}
\usepackage{fullpage}
\usepackage{natbib}

\usepackage{avant}

\usepackage{color}
\usepackage{url}
\usepackage{lineno}

\usepackage{graphicx}
\usepackage{dcolumn}
\usepackage{bm}
\usepackage{amsfonts}
\usepackage{amsmath, amssymb}

\newcommand\qed{\phantom{\underline{y}}\hfill\hfill$\square$}

\newcommand{\weg}[1]{}
\title{\scshape Is there long-range memory in solar activity on time scales shorter than the sunspot period?}

\author{M. Rypdal$^1$ and K. Rypdal$^2$ \\ ~\\ {\small $^1$Department of Mathematics and Statistics, University of Troms{\o}, Norway.} \\ {\small $^2$Department of Physics and Technology, University of Troms{\o}, Norway.}}

\begin{document}
\maketitle

\begin{abstract}
The sunspot number (SSN), the total solar irradiance (TSI), a TSI reconstruction, and the solar flare index (SFI), are analyzed for long-range persistence (LRP). Standard Hurst analysis  yields $H \approx 0.9$, which suggests strong LRP. However, solar activity time series are non-stationary due to the almost periodic 11 year smooth component, and the analysis does not give the correct $H$ for the stochastic component. Better estimates are obtained by detrended fluctuations analysis (DFA), but estimates are biased and errors are large due to the short time records. These time series can be modeled as a stochastic process of the form $x(t)=y(t)+\sigma \sqrt{y(t)}\, w_H(t)$, where $y(t)$ is the smooth component, and $w_H(t) $ is a stationary fractional noise with Hurst exponent $H$. From  ensembles of numerical solutions to the stochastic model, and application of Bayes' theorem, we can obtain bias and error bars on $H$ and also a test of the hypothesis that a process is uncorrelated ($H=1/2$). The conclusions from the present data sets are that SSN, TSI and TSI reconstruction  almost certainly are long-range persistent, but with  most probable value $H\approx 0.7$. The SFI process, however, is either very weakly persistent  ($H<0.6$) or completely uncorrelated. Some differences between stochastic properties of the TSI and its reconstruction indicate some error in the reconstruction scheme.

\end{abstract}

\section{Introduction}$\label{Intro}$
As the Sun is  the main energy source driving Earth processes, variability of Solar output has been at the center of scientific interest for centuries. Since Galileo's invention of the telescope, sunspots have been systematically counted. It has been noticed that the sunspot number (SSN) fluctuates in a more or less random fashion from month to month, but that their annual averages vary quasi-periodically with a period of approximately 11 years. Other cycles (for instance on 22 and 80 years) are also discernible in  the sunspot time series, and times of almost vanishing sunspot activity have been detected within the historical records. One example is  the Maunder minimum 1630-1720 when sunspots almost disappeared for nearly a century.

Modern telescopes, and since the late 1970s also space observatories,  have also facilitated recording the variations of solar-flare activity, for instance through the solar flare index (SFI) and measured variability of total solar irradiance (TSI) and more recently the solar spectral irradiance (SSI). Variations of TSI obviously represent a forcing of Earth climate, while the effect of variations of SSI is more subtle, since different parts of the solar spectrum affect the Earth atmosphere in different ways. Other climatic influences of solar variability are strongly debated, such as the effect of changes in the galactic cosmic ray flux due to variations in the Sun's open magnetic flux associated with the solar wind.

Much attention has been devoted to study correlation between solar activity and climate on centennial time scales through various reconstructions of solar activity based on theoretical models with sunspot data as input and by means of paleo-data on millennial scales (see review by \cite{Usoskin2008}). This interest is obvious since a positive and quantitative identification of such a connection would allow a direct empirical estimate of climate sensitivity to solar activity as reviewed by \cite{Haigh2007,Gray2010}.

The idea of long-range persistence in the solar records, on short as well as long times scales, is not new. \cite{Mandelbrot1969} applied rescaled-range ($R/S$) analysis to monthly SSN and found the characteristic bulge on the $\log (R/S)$ vs.  $\log \tau$ curve for time scales $\tau$ around the period of the sunspot cycle, but a common slope corresponding to a Hurst exponent of $H\approx 0.9$ in the ranges $3<\tau<30$ months and $30<\tau<100$ years. \cite{Ruzmaikin1994} obtained similar results for the SSN, and for analysis extended to the time range $100<\tau<3000$ years by using a $^{14}C$ proxy for cosmic ray flux. Since such behavior of the $R/S$-curve is reproduced by adding a sinusoidal oscillation to a fractional Gaussian noise with Hurst exponent $H$, both groups concluded that the stochastic component of the SSN is a strongly persistent fractional noise in the time range from months to millennia. \cite{Ogurtsov2004} combined SSN and SSN reconstructions with $^{14}C$ proxies to obtain $H$ in the range 0.9-1.0 in the time range $25<\tau<3000$ years by means of $R/S$ analysis and first and second order detrended fluctuation analysis (DFA(1) and DFA(2)) \citep{Huang}. The statistical significance of the long-memory persistence hypothesis on time scales longer than the sunspot period has been questioned by \cite{Oliver}, using the so-called scale of fluctuation approach.

Another class of approaches to the sun-climate connection has focused on the detection of  correlations and/or  common statistical signatures between proxies of solar activity and climate signals on  time scales of the sunspot period and shorter. Since the correlation between the  solar cycle variation of TSI and climate signals like global temperature anomaly (GTA) appears to be quite weak, some works have drawn the  attention to a possible ``complexity-linking'' between solar activity and Earth climate which is proposed to be discernible by identification of a common long-range memory process in the solar and climate signals, represented by a common Hurst exponent $H$ \citep{Grigolini,SW2003, SW2005,SW2008, Scafetta2004}. This view was  criticized by us in a recent Letter \citep{RR-PRL}, and triggered a comment by \cite{SWcomment} and a reply; \cite{RR-reply}. Our view is that trends, like the sunspot cycle in solar signals and multi-decadal oscillations superposed on a rising trend in global temperature signals, will create spuriously high Hurst exponents. For instance \citep{SW2005} (their Figure 3B) obtain $H\approx 0.95$ for solar as well as global temperature signals. After such trends are accounted for, both solar signals and climate signals exhibit considerably lower Hurst exponents for the stochastic component of the signals. In the present paper we will focus on three proxies for solar activity, SSN,  TSI, and SFI, and  demonstrate that the SSN and TSI  are the most persistent (most probable $H$-values 0.73 and 0.70, respectively)  on time scales shorter than the sunspot period. Further, for the SFI the most probable value is 0.54, and we cannot falsify the null hypothesis that the stochastic component in the solar flare signals exhibit no long-range memory  ($H=0.5$) on time scales 10-1000 days.

\section{Estimating Hurst exponents from data} 
\subsection{The special role of the variogram}\label{sec:variogram}
Let $x(1),x(2),x(3),\dots$ be a stationary process (a noise). 
We suppose that there exists a continuous-time motion $X(t)$, with stationary increments, such that $x(n)=X(n+1)-X(n)$. The $q$'th order structure function $S_q(t)$ is the $q$'th statistical moment
$$
S_q(t)\equiv {\mathbb E}[X(t)^q].
$$
If the motion $X(t)$ is statistically self-similar the  structure functions are power laws
$$
S_q(t)\propto t^{\zeta(q)},
$$
and  the scaling function $\zeta(q)$ has a linear dependence on $q$. 
$$
\zeta(q)=Hq,
$$
where $H$ is the self-similarity exponent. For  a self-similar motion the probability density function (PDF) of  $X(t)/\sigma (t)$, where  $\sigma^2(t)\equiv S_2(t)\propto t^{2H}$, is the same for all $t$. If this PDF is Gaussian the process is called a fractional Brownian motion (fBm). For an important class of processes, called multifractal, the PDFs change shape with varying time scale. They typically have higher flatness (kurtosis) for small $t$, often converging towards Gaussian for large $t$. For such processes the structure functions are still power laws, but the  scaling function is not linear, but has a convex shape. In this paper, we shall deal with an even wider class of processes where only the  second-order structure function, sometimes called the variogram, is assumed to be a power-law, i.e.,
\begin{equation}
S_2(t)\equiv {\mathbb E}[X(t)^2] ={\mathbb E}[X(1)^2]\,  t^{2H}. \label{vario}
\end{equation}
We call $H$ the Hurst exponent of the process $x(t)$. If $X(t)$ is a self-similar process, then $H$ is its self-similarity exponent. Thus, in general we shall not require $X(t)$ to be neither Gaussian or self-similar. In fact, we do not even require that the process is multifractal since we only assume that the second structure function is a power law.  The importance of the Hurst exponent is its relation to correlations in the noise $x(t)$. We  include the derivation of this relationship for completeness, and start by writing
\begin{equation}
2\, X(t)X(s) = X(t)^2+X(s)^2-[X(t) -X(s)]^2.  \label{correl}
\end{equation}
Since $X(t)$ has stationary increments, and using equation (\ref{vario}), we have that $${\mathbb E}[(X(t) -X(s))^2]={\mathbb E}[X(t-s)^2]={\mathbb E}\, [X(1)^2]|t-s|^{2H}$$,
and
$$
 {\mathbb E}[X(t)^2] = {\mathbb E} [X(1)^2]\, t^{2H}\, , \, \, \,  {\mathbb E}[X(s)^2] =  {\mathbb E}[X(1)^2]\, s^{2H},
 $$
 which, using equation (\ref{correl}),  yields the co-variance structure  
\begin{equation} \label{covar}
{\mathbb E}[X(t)X(s)] = \frac{1}{2} {\mathbb E}[X(1)^2]\Big{(} t^{2H}+s^{2H}-|t-s|^{2H} \Big{)}.
\end{equation}
Choosing $X(0)=0$ we now have 
\begin{eqnarray*}
{\mathbb E}[x(1)x(n)] &=& {\mathbb E}\Big{[}\Big{(}X(1)-X(0) \Big{)}  \Big{(}X(n+1)-X(n) \Big{)} \Big{]} \\Ê
&=& {\mathbb E}[X(1)X(n+1)]- {\mathbb E}[X(1)X(n)] ,
\end{eqnarray*}
and by using  equation (\ref{covar})we  get
\begin{eqnarray*}
{\mathbb E}[x(1)x(n)]  &\propto& (n+1)^{2H} -2n^{2H}+(n-1)^{2H}  \\Ê
&\sim& \frac{d^2}{dn^2} n^{2H} = 2H(2H-1)\,n^{2H-2}.
\end{eqnarray*}
This means that the correlation function of $x(t)$ has algebraic decay for all $H \in (0,1)$ except for $H=1/2$, for which $x(t)$ is an uncorrelated noise.  For $H>1/2$ the integral over the correlation function is infinite, and this property is what is denoted as long-range persistence. Again, we emphasize that this property only requires a power-law dependence of the variogram, but not that the process is self-similar or even multifractal.
\subsection{Detrended fluctuation analysis} \label{sec:DFA}
As mentioned in the introduction solar signals contain distinct periodicities, the most prominent being the 11-year solar cycle, and these trends will distort the result of the variogram analysis. In the next section we will demonstrate several ways of detrending and how these tend to reduce the estimated Hurst exponents. Here we shall just briefly describe one systematic method, the detrended fluctuation analysis (DFA), which performs quite well on these data. Again  we consider the discrete and stationary stochastic process $\{x(k)\}$ and we construct the cumulative sum process $X(i)=\sum_{k=0}^i x(k)$. Let us divide the sequence $X(i)$ into $N_s$ segments of length $\Delta t$, where each segment is indexed by $l=1,2,\ldots, N_s$. Then compute an $n$'th order polynomial fit $p_{\Delta t, l}^{(n)}$ to $X(i)$ in each segment and compute the variance of the detrended fluctuation in the segment,
$$
V_{\Delta t, l}^{(n)}\equiv \frac{1}{\Delta t}\sum_{i=1}^{\Delta t}[X_{\Delta t, l}(i)-p_{\Delta t, l}^{(n)}]^2.
$$
The square root of the average of this variance over all the segments,
$$
F^{(n)}(\Delta t)\equiv \left( \frac{1}{N_s}\sum_{l=1}^{N_s}V_{\Delta t, l}^{(n)}\right)^{1/2}
$$
is the $n$'th order DFA fluctuation function. If an fBm with Hurst exponent $H\approx 0.5$ is superposed on a sinusoidal signal of comparable amplitude, ordinary variogram analysis will give an estimated Hurst exponent $H\approx 1$, while a third order DFA analysis (DFA(3)) will give
$$
F^{(n)}(\Delta t)\propto \Delta t^{H}
$$
with $H$ close to the the Hurst exponent for the fBm. Thus, there are good reasons to assume that given sufficiently long data series DFA(3) would give an accurate estimate of the Hurst exponent for the underlying detrended stochastic process. The main problem addressed in this paper is how to reduce the uncertainties in the  estimates that arise from the limited data sets. But first  we demonstrate the  effect of detrending on the solar data without dealing with uncertainties.
\subsection{The effect of trends}

\begin{figure}[t]
\begin{center}
\includegraphics[width=7.5cm]{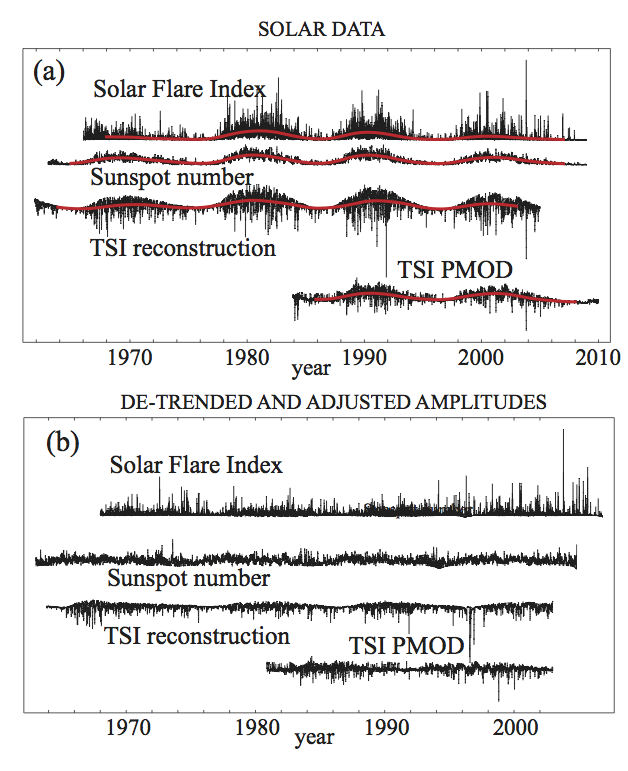}
\caption{
(a): Time series for Sunspot Number (SN), Solar Flare Index (SFI),  Total Solar Irradiance (TSI) reconstruction, and the TSI PMOD composite. The smooth curves are obtained by convolving the time series with a gaussian kernel with a one-year standard deviation. The data are normalized such that the smooth components have equal maximal amplitude, i.e. the difference between the maximal and minimal value is the same for the four smoothed signals. (b): In this figure we have subtracted the smoothed signals from the original times series in order to de-trend the time series and then divided the detrended signals by the square root of the smoothed signals in (a).  Prior to this amplitude adjustment, the origins for each of the signals are shifted to the minimal values of the smooth signals in (a). 
} \label{fig1}
\end{center}
\end{figure}

\begin{figure}[t]
\begin{center}
\includegraphics[width=7.0cm]{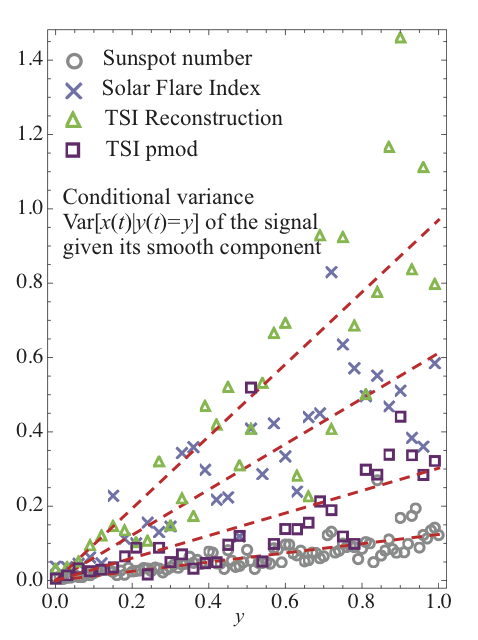}
\caption{
For each of the signals $x(t)$ in Figure \ref{fig1}a the plots show the variance $\mbox{Var}[x(t)|y(t)=y]$ conditioned upon the value of the corresponding smoothed signal $y(t)$. When the variance's dependence on the smooth signal of the fluctuations are well approximated by a linear function, we can model the standard deviation's dependence on the smooth signal by the square-root function.  
} \label{fig2}
\end{center}
\end{figure}

In Figure \ref{fig1}a we have plotted the SFI, SSN and a TSI reconstruction over the last four sunspot cycles, and the instrumental TSI PMOD composite for the last two cycles. All data are given as daily averages. The thick smooth curves are moving averages weighted by a Gaussian window with 1 year standard deviation.  In order to extract the stochastic component the smoothed signal should be subtracted, but the signals are still strongly non-stationary because the amplitudes of the rapid fluctuations  are much higher at sunspot maxima than at minima. These amplitudes turn out to have a distinct statistical dependence on the local moving average $y(t)$. In fact, the variance of the fluctuations around $y(t)$ is roughly proportional to $y(t)$, i.e. $\mbox{Var}[x(t)|y(t)=y]\propto y$, as shown in Figure \ref{fig2}.

\begin{figure}
\begin{center}
\includegraphics[width=7cm]{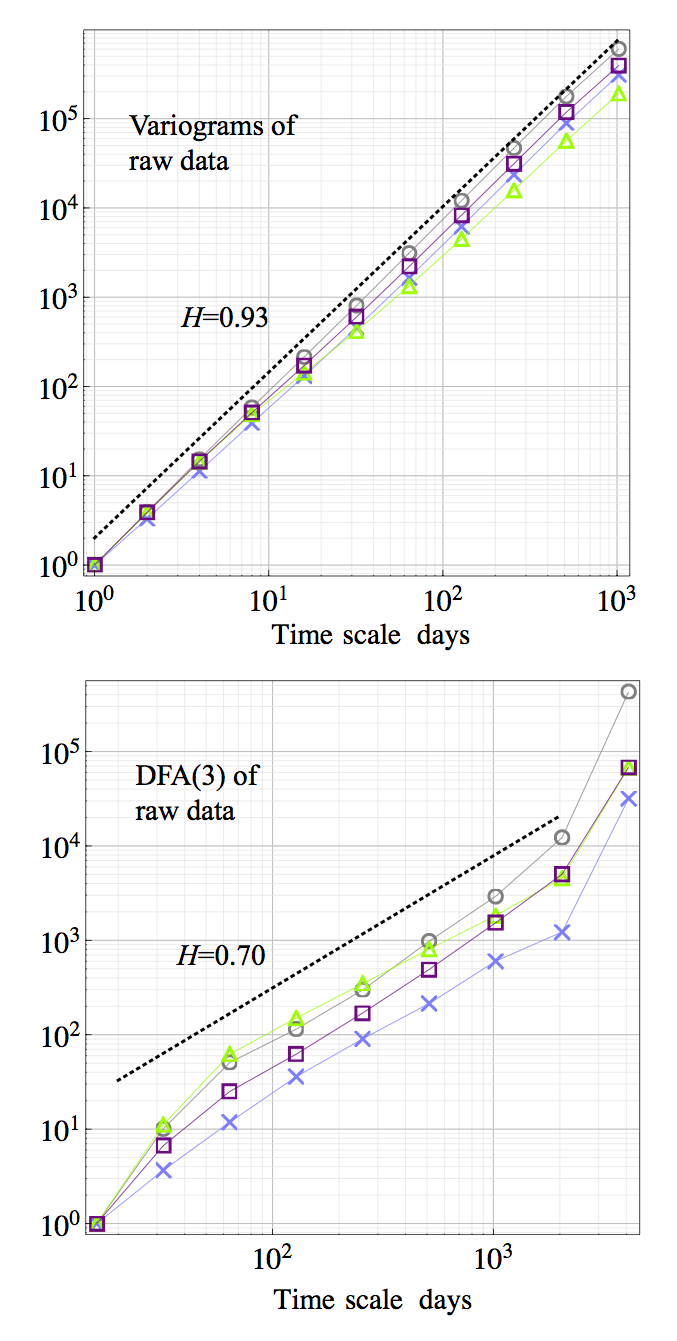}
\caption{Variograms computed from the raw (undetrended) data. (b): DFA(3) variogram (square of fluctuation function) computed from raw data. The values for $H$ given in the figure is half the slope of the dotted lines. These lines are drawn to indicate the typical slope of the corresponding variograms.} \label{fig3}
\end{center}
\end{figure}

 This means that the mean- and amplitude-detrended signals 
 \begin{equation}
 z(t)\equiv (x(t)-y(t))/\sqrt{y(t)}, \label{detrended}
 \end{equation}
which is plotted in Figure \ref{fig1}b, are approximately stationary. The sunspot minima  are still discernible in the detrended signals since flares and sunspots are absent for long time intervals around these minima. This implies that if statistics  for the active sun is sought, it may be  reasonable  to eliminate periods of very low activity around sunspot minima from the analysis. In Figure \ref{fig3}a we have plotted variograms of the four raw signals $x(t)$. This yields Hurst exponents in the range $0.88<H<0.97$ and corresponds to the results obtained in Figure 3B of \cite{SW2005}. In Figure \ref{fig3}b we show results of DFA(3) analysis, which yields Hurst exponents in the range $0.55<H<0.67$. Results similar to the latter is obtained by computing variograms of the detrended signal in  Figure \ref{fig1}b.

\section{A stochastic model}

As mentioned in section \ref{sec:DFA} the limited length of the observed data records makes it  difficult to compute error bars in the estimates of Hurst exponents directly  from the data. Essentially the entire data set is used to estimate one value, and then we do not have more data to assess the statistical spread in this estimate. What we need to know to compute such error bars is the PDF of estimated values $\hat{H}$ in an imaginary ensemble of realizations of data sets of the same length as the observed record. Such an ensemble can be generated synthetically from a model that is assumed to have the the same statistical properties as the observed data, including an hypothesized value of $H$ that can be varied. From the numerically generated ensembles one can construct a  conditional probability  density $p(\hat{H}|H)$ for obtaining an estimated value $\hat{H}$ for the Hurst exponent, given that the ``true" exponent is $H$. Then, by means of Bayes' theorem we can obtain the conditional PDF $p(H|\hat{H})$, which gives us the probability of having a ``true" Hurst  exponent $H$ provided we have estimated a value $\hat{H}$ from the observational data. The width of $p(H|\hat{H})$ gives us the error bar of our  estimate. The details of this procedure will be given in the next section, but first we shall describe the model used to generate synthetic data, which follows naturally  from equation (\ref{detrended}). We have  already observed that $z(t)$ defined from this expression is an approximately stationary stochastic process and that variograms estimated from the data are  approximate power laws, from which a Hurst exponent can be estimated. PDFs of the signals on  different time scales turn out to have approximately the same shape on the different time scales, but they are not always Gaussian. We shall therefore model $z(t)$ as a self-similar (in general non-Gaussian) process with self-similarity exponent $H$. By denoting this fractional noise process by $w_H(t)$, we can write
\begin{equation} \label{model}
x(t) = y(t) + \sigma\,\sqrt{y(t)}\,w_H(t),
\end{equation}
where $y(t)$ is a 11-year-periodic oscillation with amplitude $A$, and $w_H(t)$ is a fractional non-Gaussian noise with unit variance. The relation between parameters $\sigma$ and $A$ are easily estimated from the sunspot data by comparing the amplitude of the smoothed signal in Figure \ref{fig1}a with the amplitude of the detrended noise in Figure \ref{fig1}b.

The fractional noise $w_H(t)$ is generated by taking $w_H(t)=Z_H(t+1)-Z_H(t)$ where $Z_H(t)$ is a process defined by the integral 
\begin{equation} \label{modeldef}
Z_H(t) = c_H \int K_H(t,t')\,dZ(t')\,.
\end{equation}
Here $c_H$ is a normalization constant and 
$$
K_H(t,t') =(t-t')_+^{H-1/2}-(-t')_+^{H-1/2} \,.
$$
The process $Z(t)$ is chosen to be a L{\'e}vy process (a process with stationary and independent increments) with zero mean and unit variance. If $Z(t)$ is gaussian, implying that it is a Brownian motion, then $Z_H(t)$ is a fractional Brownian motion with self-similarity exponent $H$. In this case $w_H(t)$ is a fractional Gaussian noise with Hurst exponent $H$. 
Using equation (\ref{modeldef}) together with the independence of increments in $Z(t)$ we obtain the relation 
$$
E[Z_H(t)^2] =c_H^2 \int K_H(t,t')^2\,\,dt'\,, 
$$
and using the scaling relation $K_H(at,at')=a^{H-1/2}K_H(t,t')$ it follows that $E[Z_H(t)^2] \propto t^{2H}$. 

For any given distribution\footnote{In order for $Z(t)$ to be a well-defined L{\'e}vy process these distributions must be infinitely divisible. However, since we do not require our model to hold on time scales shorter than a day, this has no practical importance.} on $Z(t)$ (and hence on $w_H(t)$) we can simulate the process $Z_H(t)$ using the method of \cite{Stoev2004}. For the Sunspot Number the estimated PDF of the noise $w_H(t)$ is shown in figure \ref{fig6}(b). We see that it is reasonable to choose a Gaussian distribution for $w_H(t)$. For the other time series the Gaussian approximation is not valid (See figures \ref{fig4}(b) and \ref{fig10}(b)), and here we fit the data to stable distributions using a maximum likelihood estimator. These stable distributions are then truncated at values corresponding roughly to the largest observed datapoint in the time series. The truncation ensures that the moments of $w_H(t)$ exist and prevents the presence of unrealistically large events that can influence the analysis of estimated Hurst exponents. To demonstrate that our estimated distributions are reasonable we have compared the estimated PDFs for the fluctuations in the solar data with PDFs estimated from realizations of the stochastic models. These are presented in figures \ref{fig4}(b), \ref{fig6}b and \ref{fig10}(b).     

\section{A Bayesian estimate}
\subsection{Uniform sample space}
We assume that the observed signal is a realization of the stochastic equation (\ref{model}) where $w_H(t)$ is a fractional noise with the measured distribution and with Hurst exponent $H\in [0,1]$. We construct a uniform (all values of $H\in I\equiv [0,1]$ are equally probable) sample space  $\cal S$ of realizations of these processes. For each realization with a given $H$ we measure (estimate) a value $\hat{H}$. In general $\hat{H} \neq H$.

We can then, for instance, compute the conditional probability density  $p(\hat{H}|H)$ from an ensemble of numerical solutions to the stochastic model where for each realization $H$ is chosen randomly in the interval $I$. The conditional PDF $p(H|\hat{H})$ can  also be computed directly from the ensemble or from Bayes' theorem:
\begin{equation}
p(H|\hat{H})=\frac{p(\hat{H}|H)p(H)}{p(\hat{H})}. \label{Bayes1}
\end{equation}
Here $p(H)$ is by construction of $\cal S$ uniform on the unit interval $I$ and hence $p(H)=1$. However, $p(\hat{H})$ is not necessarily uniform and must be computed from the ensemble.
\subsection{A non-uniform sample space}\label{nonuniform}
The uniform sample space is based on a model of reality {\em prior to any observation} of $\hat{H}$ that any value of  $H\in I$ is equally probable, i.e. we have no prior preference to what $H$ should be. This may not be the most plausible starting point. Another is to ask the question; is the stochastic component in the signal long-range correlated or not? In other words, is $H=1/2$ or is $H\neq 1/2$? In the uniform sample space we have of course that $p(H=1/2)=0$, so this sample space is not a good starting point if we want to assign a non-zero prior probability to the hypothesis that $H=1/2$. What we can do is to construct a non-uniform sample space ${\cal S}(p_0)$ consisting of an ensemble of numerical realizations of the fractional process where a fraction $p_0$ of the realizations have $H=1/2$ and the remaining fraction $1-p_0$ is chosen with $H$ drawn randomly from the set $I_{H\neq1/2}= [0,1/2)\bigcup (1/2,1]$. Let us formulate the following hypothesis: \newline
 
 \noindent
{\bf Hypothesis }$ \cal{H}$: The observed signal is described by the stochastic equation with an uncorrelated ($H=1/2$) stochastic noise term. \newline

\noindent We then define an observation: \newline

\noindent
{\bf Observation} $H_o$: DFA(3) applied to the observed signal yields an estimated Hurst exponent $\hat{H}=H_o$.\newline

\noindent From the constructed nonuniform ensemble we can  compute the conditional probability that the hypotesis $\cal H$ is true provided the prior probability of this being true is $p_0\equiv p(\mathcal{H})$ and we have made the observation by DFA of the physical time series that the estimated Hurst exponent is $\hat{H}=H_o$. 

First we must create an ensemble ${\cal S}$ with $N$ elements. $N p_0$ elements are realizations with $H=1/2$. We call this subensemble ${\cal S}_{H=1/2}$. The remaining $N (1-p_0)$ elements are realizations  with $H$ drawn randomly from $I_{H\neq 1/2}$. We call this subensemble ${\cal S}_{H\neq 1/2}$. By doing this we have implicitly assumed zero probability to have $H$ outside the unit interval, and hence that $p(\mathcal{H})+p(\bar{\mathcal H})=p_0+(1-p_0)=1$. Here, the symbol  $\bar{\mathcal H}$ denotes the null hypothesis, $H\in I_{H\neq 1/2}$.  The conditioned subensemble ${\cal S}(H_o)$ is the set of all elements for which the estimated $\hat{H}$ is in an $\epsilon$-neighborhood of the observed value $H_{o}$. Here $\epsilon$ should be a small number, but not smaller than the measurement error.  Let us consider  the fraction $p(H_o|{\mathcal H})$ of elements in ${\cal S}_{H=1/2}$ which belong to  ${\cal S}(H_o)$ and the fraction $p(H_o|{\bar{\mathcal H}})$ of ${\cal S}_{H\neq 1/2}$ which belong to ${\cal S}(H_o)$. These fractions are the conditional probabilities that $\hat{H}\in (H_o-\epsilon,H_o+\epsilon)$ provided the hypotheses $\mathcal{H}$ or  the  hypothesis ${\bar{\mathcal H}}$ are true, respectively. It is now convenient to introduce  the Bayes factor:
\begin{equation}
{\cal R}\equiv\frac{p(H_o|{\mathcal H})}{p(H_o|\bar{\mathcal H})}. \label{Bayesfactor}
\end{equation}
Note that if $\epsilon$ is small, both numerator and denominator in this ratio are proportional to $\epsilon$, and hence $\cal R$ is independent of $\epsilon$. The Bayes factor is large (${\cal R}\gg 1$) if the probability of making the observation $H=H_o$ is much larger if the hypothesis $\mathcal{H}$ is true than if $\bar{\mathcal H}$ is true. Intuitively then, making the observation $H_o$ should support the hypothesis and make $p({\mathcal H}|H_o)$ close to unity. On the other hand, we have that ${\cal R}\ll 1$ if the probability of  $H=H_o$ is much larger if $\bar{\mathcal H}$ is true than if  $\mathcal{H}$ is true. In that case one should intuitively expect $p({\mathcal H}|H_o)$ to be vanishingly small. This can be quantified by means of 
 Bayes' theorem, which yields, 
$$
\frac{p({\mathcal H}|H_o)}{p(\bar{\mathcal H}|H_o)} = {\cal R}\,  \frac{p({\mathcal H})}{p(\bar{\mathcal H})} = {\cal R}\,  \frac{p_0}{1-p_0}\,.
$$
Since $p(\bar{\mathcal H}|H_o)=1-p({\mathcal H}|H_o)$ we can insert this in the equation above and solve for $p({\mathcal H}|H_o)$: 
\begin{equation}
p({\mathcal H}|H_o)=\left(1+\frac{1}{\cal R} \, \frac{1-p_0}{p_0}\right)^{-1}. \label{Bayesresult}
\end{equation}
We observe that in addition to the Bayes factor also the prior probabilty $p_0$ enters the expression, but the result is not extremely  sensitive to the choice of $p_0$ provided it is not very close to 0 or 1. The prior probability  seeks to quantify our prior knowledge about the truth of the hypothesis $\mathcal{H}$, and if that knowledge is poor, the most reasonable choice may be $p_0=1/2$, which makes $(1-p_0)/p_0=1$ and
$$
p({\mathcal H}|H_o)=\frac{\cal R}{1+\cal R}.
$$
 In general, the conditional probabilities entering the Bayes factor may be computed from empirical data sets or from theoretical models. In our case, we are facing very limited data sets, and we therefore compute them from an ensemble of  numerically generated realizations of the stochastic  equation (\ref{model}).
The estimated Bayes factors for the various solar proxies are presented in Table 1.

\section{Results}
Since we are making Bayesian estimates, the results depend on our prior probablities. Using the uniform ensemble, this probability distribution is uniform on the unit interval. Using the non-uniform sample space the prior probability parameter is $p_0$.

\subsection{The data}
The data records  analyzed in the following are selected  mainly for illustration of the power and limitations of the analysis methods, although they also represent interesting proxies that reflect different aspects of solar activity. All records analyzed  have different lengths, the longest is the sunspot number, and the shortest is the TSI composite. It will become clear that the error bars in the estimates of the Hurst exponents depend on the lengths of the records, which is why the smallest error bars are found in the sunspot number and the largest in the TSI composite. It is conceivable that other existing data sets could be used to reduce these error bars, and  certainly the near future will provide such data sets.

The solar flare index (SFI) data have been downloaded from the NOAA National Geophysical Data Center (NNDGC) in Boulder, Colorado. The file contains daily data from the beginning of 1966 to the end of 2008. There are no missing data points in this record. There is certainly a wealth of other solar-flare data available, but to our knowledge this is the longest unbroken time-record of solar flare activity with daily resolution.

The daily sunspot number has been downloaded from the Solar Influences  Data Analysis Center (SIDC) in Brussels, Belgium. The file contains data from 1818, but have missing data points up to 1848. After this year the data record is complete, so in our analysis we have used the unbroken record from 1848 to 2011.

The TSI reconstruction is based on \cite{Krivova2007} and the data file is retrieved from Max-Planck Institute of Solar System Research (MPS) in Lindau am Hartz, Germany. The data file contains daily data from 1611, but with many gaps of missing data points before 1963. For our analysis we therefore use the unbroken record from 1963 to the end of 2004.

The TSI   composite data record is based on \cite{Frolich2000} and is downloaded from  Physikalisches-Meteorologisches Observatorium Davos/World Radiation Center (PMOD/WRC), Switzerland, and is version d41\_62\_1007. The daily data set goes from January 1976 to July 2010, but contains 7\% missing data points. The missing data dominates in the early part of the record,  and for this reason we have discarded all data before 1983. For the data after 1983 there are only a small fraction of the data points missing. These missing data points have been reconstructed from the stochastic model equation (\ref{model}), using a non-Gaussian white noise for $w_h(t)$. The probability density function (PDF) for this noise term is estimated from the data, and the drift term $y(t)$ is chosen to be the smoothed trend curve shown in figure \ref{fig2}. Since this smoothed trend curve is based on a weighted moving average $y(t)$ is properly defined only up to two years after the start and two years before the end the data record, so the time series analyzed in the following goes from 1985 to 2008.

\subsection{Uniform sample space}
\begin{figure}
\begin{center}
\includegraphics[width=8cm]{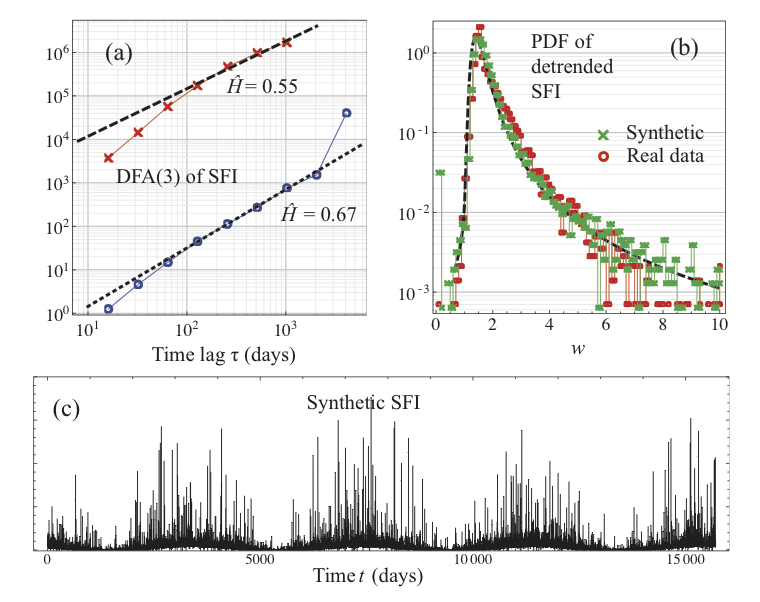}
\caption{(a): DFA(3) fluctuation function squared computed from the SFI raw data (circles), and DFA(3) fluctuation function computed from the SFI raw data, but by omitting all intervals on all time scales that contain data points where the SFI is zero. (b): PDF of fluctuations  the detrended SFI data series (circles), and PDF of the corresponding data generated by the stochastic model. (c):  A realization of a model-generated SFI time series.}
 \label{fig4}
\end{center}
\end{figure}

Figure \ref{fig4}a shows the DFA(3) fluctuation function for the observed SFI time series (lower curve) in a log-log plot. The estimated Hurst  exponent is $\hat{H}\approx 0.67$. For the SFI the DFA(3) estimate is particularly sensitive to the solar minimum periods because in these intervals there are long intervals with no solar flares and zero SFI, which gives rise to spurious persistence of the signal. If intervals where the index is zero is excluded from the analysis on all time scales, we obtain the upper curve in the figure, which yields $\hat{H}\approx 0.55$. We belive this figure is more representative for the solar-flare statistics in the active sun. Figure \ref{fig4}b shows the PDF of the  detrended and amplitude-adjusted SFI (shown in Figure \ref{fig1}c) together with the PDF of the synthetically generated SFI-signal. The dashed curve is a fit of a stable distribution to the estimated PDF. The lower panel Figure  \ref{fig4}c shows a realization of the  synthetically generated un-detrended signal. This signal should be compared to the upper trace in Figure \ref{fig1}a.

\begin{figure}
\begin{center}
\includegraphics[width=7.5cm]{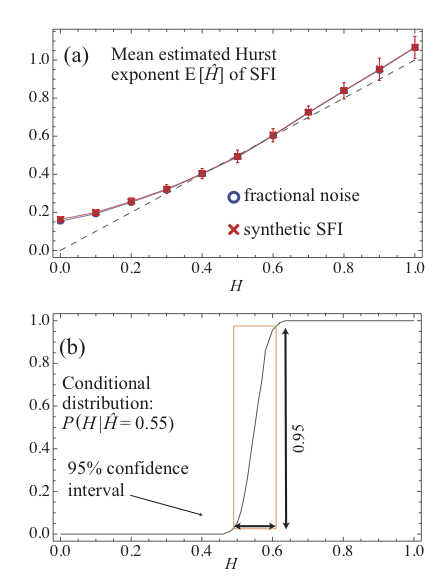}
\caption{
(a): Ensemble mean $E[\hat{H}]$ of estimated Hurst exponent for SFI computed by the DFA(3) method. The crosses are computed from an ensemble of realizations generated  from the stochastic model where  the fractional noise $w_H(t)$ used in the  model has Hurst exponent $H$ and the PDF given in Figure \ref{fig4}b. The circles (partly hidden by the crosses) are computed from DFA(3) applied to the fractional noise directly.  The Gaussian statistical spread in the estimated $\hat{H}$ is given by the error bars. (b): The conditional cumulative distribution $P(H|\hat{H})\equiv \int_{-\infty}^{H}p(H'|\hat{H})dH'$ computed from  the conditional PDF $p(\hat{H}|H)$ by means of the uniform sample space and Bayes' theorem for $\hat{H}=0.55$.  } \label{fig5}
\end{center}
\end{figure}

\begin{figure}
\begin{center}
\includegraphics[width=8.5cm]{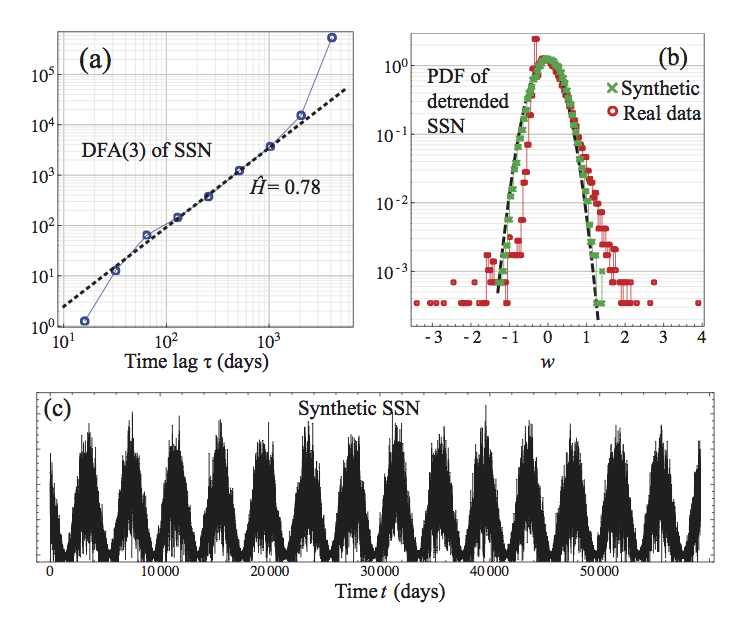}
\caption{
(a): DFA(3) fluctuation function squared computed from the SNN raw data (circles). (b): PDF of fluctuations  the detrended SSN data series (circles), and PDF of the corresponding data generated by the stochastic model. (c):  A realization of a model-generated SSN time series.
} \label{fig6}
\end{center}
\end{figure}

In Figure \ref{fig5}a (crosses) we show the mean estimated $\hat{H}$ computed from ensembles of synthetic realizations of the SFI for 11 different values of $H$ in the unit interval. $\hat{H}$ is estimated from DFA(3) applied to the synthetic $x(t)$ (i.e. the signal with the periodic trend) and the deviation from the straight dashed line indicates the systematic bias of the DFA(3) method applied to such a signal. When the process is persistent ($H>0.5$) the method performs quite well, with a slight overestimation of $H$ when the process is strongly persistent ($H\rightarrow 1$). The circles (partly hidden by the crosses) are computed from DFA(3) applied to the fractional noise directly. The fact that the two curves fall on top of each other shows that the DFA(3) method removes the influence of the periodic trend. There is also a statistical spread in the estimated $\hat{H}$ indicated by the error bars, which has a Gaussian conditional PDF $p(\hat{H}|H)$. This statistical spread is very important in our analysis, since it gives us the opportunity to use Bayes' theorem to compute the error bars on the observed Hurst exponent. The spread in estimated $\hat{H}$ depends on the length of the synthetic data record, which we have chosen equal to the observational record. As explained in the previous section the conditional PDF $p(H|\hat{H})$ can be computed from the same ensemble. In Figure \ref{fig5}b  the conditional cumulative distribution $P(H|\hat{H})\equiv \int_{-\infty}^{H}p(H'|\hat{H})dH'$ is computed from  the conditional PDF $p(\hat{H}|H)$ by means of the uniform sample space and Bayes' theorem for $\hat{H}=0.55$. From this distribution it is easy to compute  the  conditional mean ${\mathbb E}(H|H_o)$ (which is the best estimate for $H$ given the observation $H_o$) and the 95\% confidence interval for this estimate. These are given in Table \ref{tab1}, and does not rule out the possibility that the SFI stochastic process is uncorrelated ($H=0.5$). Note that ${\mathbb E}(H|H_o=0.55)=0.54$, i.e. that the best estimate for $H$ and the observation $H_o$ are slightly different. Below, we shall see that this difference is more  substantial in the sunspot number.

\begin{figure}
\begin{center}
\includegraphics[width=7.5cm]{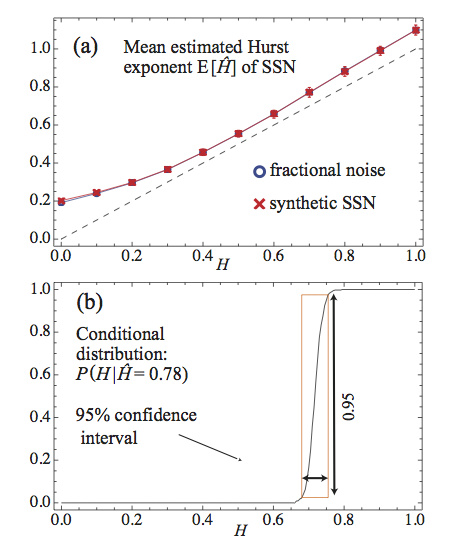}
\caption{
Same as Figure \ref{fig5}, but for SSN time series. The reason for the  narrower confidence interval in panel (b) compared to those in Figure  \ref{fig5} is that the SSN time series analyzed is considerably longer than the SFI time  series.
} \label{fig7}
\end{center}
\end{figure}

Figures \ref{fig6} and \ref{fig7} show the same for the sunspot number. Here the PDFs are nearly Gaussian, and the 95\% confidence interval is narrower than for the SFI because the observed sunspot record is much longer. The best estimate for $H$ is 0.73, while $H_o=0.78$, but the narrow confidence interval excludes the possibility that $H=0.5$.

\begin{figure}
\begin{center}
\includegraphics[width=8cm]{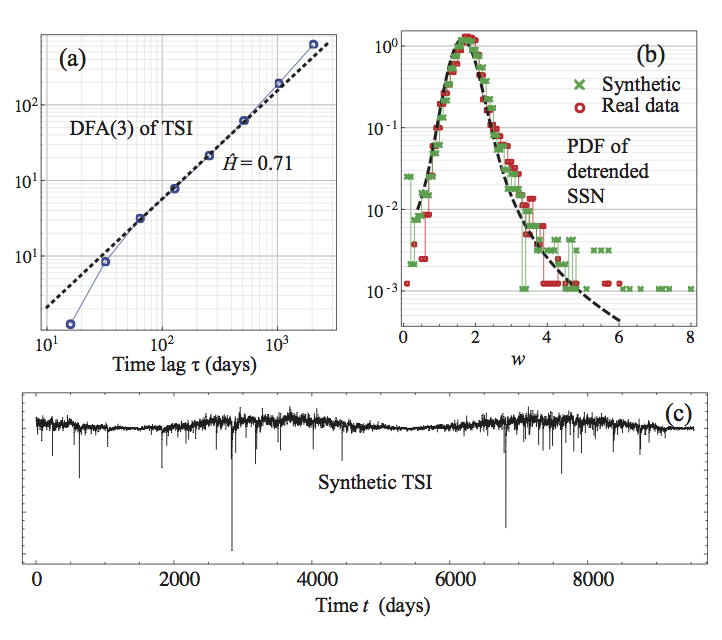}
\caption{
DFA(3) fluctuation function squared computed from theTSI raw data (circles). (b): PDF of fluctuations  the detrended TSI data series (circles), and PDF of the corresponding data generated by the stochastic model. (c):  A realization of a model-generated TSI time series.
} \label{fig8}
\end{center}
\end{figure}

\begin{figure}
\begin{center}
\includegraphics[width=8cm]{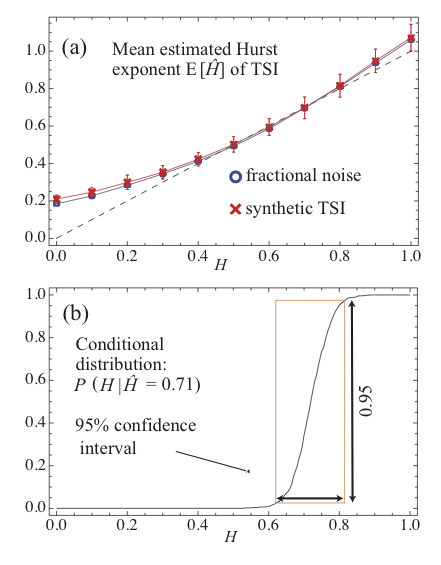}
\caption{
Same as Figure \ref{fig5} and Figure \ref{fig7}, but for the TSI time series. The reason for the wider confidence interval in panel (b) compared to those in Figure  \ref{fig5} and  \ref{fig7} is that the TSI time series analyzed is considerably shorter than the SFI and SSN time  series.
} \label{fig9}
\end{center}
\end{figure}

\begin{figure}
\begin{center}
\includegraphics[width=8cm]{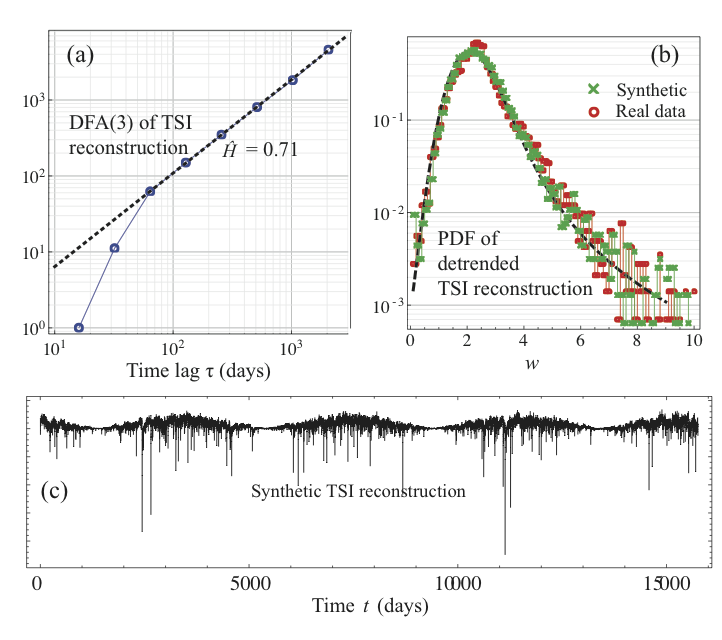}
\caption{
DFA(3) fluctuation function squared computed from the TSI reconstruction raw data (circles). (b): PDF of fluctuations  the detrended TSI reconstruction data series (circles), and PDF of the corresponding data generated by the stochastic model. (c):  A realization of a model-generated TSI reconstruction  time series. 
} \label{fig10}
\end{center}
\end{figure}

\begin{figure}
\begin{center}
\includegraphics[width=8cm]{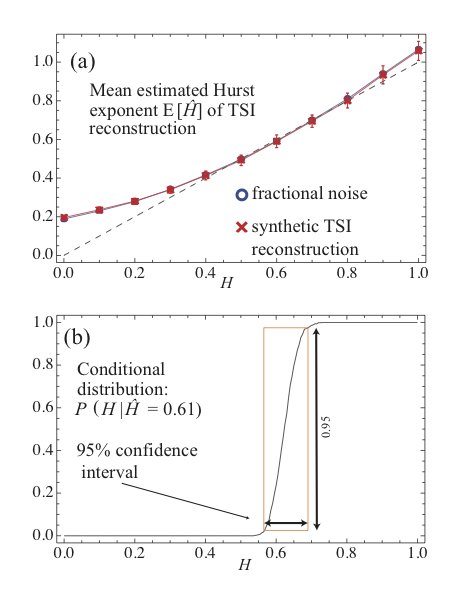}
\caption{
Same as Figure \ref{fig9}, but for the TSI reconstruction  time series. The reason for the slightly  narrower confidence interval in panel (b) compared that in  Figure  \ref{fig9} is that the TSI reconstruction time series analyzed is somewhat longer than the TSI time  series.
} \label{fig11}
\end{center}
\end{figure}

The results for the TSI are shown in Figures \ref{fig8} and \ref{fig9} and for the TSI reconstruction in Figures \ref{fig10} and \ref{fig11}. The estimated Hurst exponent is somewhat smaller ($H=0.62$) in the reconstruction  compared to the observational data ($H=0.70$). The PDF also appears more skew in the former. These differences suggests some systematic errors in the reconstruction procedure that has effect on these timescales shorter than the sunspot period. The reason for the  larger confidence intervals for the TSI compared to the reconstruction, is that the latter time series is longer.

The mean estimated Hurst exponents and their 95\%  confidence intervals are summarized in Table 1. Except for the SFI this analysis indicates that there is a significant persistence in all solar activity proxies analyzed, but their Hurst exponents are considerably lower  than obtained by most previous authors. The largest $H$ is found in the sunspot number, with the most probable value $H=0.73$, and with 95\% confidence $H<0.76$.

\subsection{Non-uniform sample space}
\begin{figure}
\begin{center}
\includegraphics[width=8.0cm]{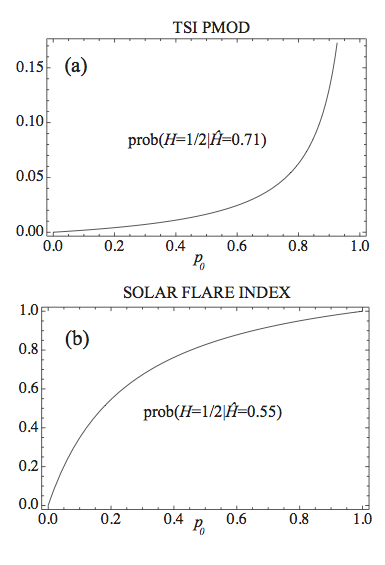}
\caption{The conditional probability $p({\cal H}| H_o)$ versus prior probability $p_0$ for  (a):TSI, and (b): SFI, computed from the non-uniform sample space.
} \label{fig12}
\end{center}
\end{figure}

From Table 1 it is tempting to conclude that $H>1/2$ also for the SFI, since the $H=1/2$ value is barely within the 95\% confidence interval. However, this result is based on an assumption on a uniform prior probability,  i.e. we assume all values of $H\in I$ is equally probable prior to the observation. If, however, prior to observation we assume that there is a finite probability that the SFI-process is uncorrelated ( $p_0\neq 0$) on the relevant time scales, section \ref{nonuniform} shows that the computation is different and the result will depend on $p_0$. The conditional probability $p({\mathcal H}|H_o)$ is given by equation(\ref{Bayesresult}), where the Bayes factor $\cal R$ must be computed from the numerically generated  ensemble of realizations generated for given values of $p_0$. The curves of $p({\mathcal H}|H_o)$ as a function of $p_0$ are shown for  TSI  and SFI in Figure \ref{fig12}. It appears that the conditional probability that $H=1/2$ is negligible for the TSI , unless we assume a prior probability for this very close to unity (in that case our prior prejudice overshadows any computation). However, for the SFI we observe that even an agnostic prior view ($p_0=0.5$) will result in a higher probability after observation $p({\mathcal H}|H_o)\approx 0.8$ and that $p({\mathcal H}|H_o)> 0.8$ if $p_0>0.5$.

\section{Discussion and conclusions}
The emphasis in this paper is on analysis method for discerning  long-range memory in short  data records with near-periodic trends in local mean and variance. The data records analyzed have been chosen mainly for illustration of these  methods. Still, from these data we can conclude that Hurst exponents for all proxies analyzed are significantly smaller than obtained recently  by \cite{SW2005} and in the classical papers by \cite{Ruzmaikin1994} and \cite{Mandelbrot1969}. We can also conclude that TSI and SSN time  series on the one hand are considerably more persistent than the SFI time series, and at present it cannot be excluded that the solar flare activity is an uncorrelated process on time scales longer than a few months. We also have indications that the particular TSI reconstruction analyzed here (see Table 1) exhibits lower persistence and a skewer PDF of the fluctuations than the observed TSI. More accurate and longer reconstruction time series are emerging, and as the TSI time series are also growing longer, the methods devised here should be able to pinpoint differences in the stochastic properties of the observed and reconstructed signals that might have to be addressed by the groups working with reconstructions.

As  measurements of  solar spectral irradiance (SSI) over several solar cycles become available, time series of irradiance in the UV-range will be interesting to analyze. Persistence in  TSI and SSI will be interesting to compare to corresponding properties in solar wind parameters, geomagnetic indices and  terrestrial climate parameters in order to assess the existence of possible sun-climate coupling \citep{SW2003,RR-PRL}, and to identify the coupling mechanisms. 

Stochastic properties of the proxy data at solar maxima could be compared with those at solar minima, and if a systematic difference exists, such measures of reconstructed data near the Maunder and Dalton minima should be compared to those away from these minima. The objective of this exercise would be to establish whether stochastic measures of proxy data could be used as reliable proxies for solar activity.



\newpage

\noindent

\noindent

\noindent

\noindent

\noindent

\noindent

\noindent


\noindent

\begin{table*} \label{tab1}
\begin{tabular}{||l|c|c|c|c||} \hline
~Ê                                      & TSI reconstruction             &TSI (PMOD)                         & Sunspot Number                   & Solar Flare Index                  \\ \hline \hline
$H_o$                         & $0.61$                                & $0.71$                               &$0.78$                                    & $0.55$    \\ \hline
 $\int H\,  p(H|H_o)\, dH$                & $0.62$                                & $0.70$                               &$0.73$                                    & $0.54$    \\ \hline
95\% confidence             & $0.57<H<0.69$                  & $0.62<H<0.81$                 & $0.68<H<0.76$                      & $0.49<H<0.61$                   \\ \hline 
$\cal R$            & $7.0 \times 10^{-3}$           & $1.6 \times 10^{-2}$                            &~                                    & $3.8$    \\ \hline

\end{tabular}

\caption{1. row: Hurst exponent $H_o$ as estimated by DFA(3) from the observed time series. 2. row: The most probable Hurst exponents $\int H\,  p(H|H_o)\, dH$ computed from the uniform sample space. 3. row: The 95\% confidence intervals computed from the uniform sample space. 4. row: The Bayes factor computed from the non-uniform sample space. The computaion for the SSN is not shown since Bayes factor in this case is obviously extremely small.}
\end{table*}
\end{document}